# Analysis of Spent Nuclear Fuel Imaging Using Multiple Coulomb Scattering of Cosmic Muons

S. Chatzidakis*, C. K. Choi, L. H. Tsoukalas

Purdue University, School of Nuclear Engineering, West Lafayette, IN 47907

Cosmic ray muons passing through matter lose energy from inelastic collisions with electrons and are deflected from nuclei due to multiple Coulomb scattering. The strong dependence of scattering on atomic number Z and the recent developments on position sensitive muon detectors indicate that multiple Coulomb scattering could be an excellent candidate for spent nuclear fuel imaging. Muons present significant advantages over existing monitoring and imaging techniques and can play a central role in monitoring nuclear waste and spent nuclear fuel stored in dense well shielded containers. The main purpose of this paper is to investigate the applicability of multiple Coulomb scattering for imaging of spent nuclear fuel dry casks stored within vertical and horizontal commercial storage dry casks. Calculations of muon scattering were performed for various scenarios, including vertical and horizontal fully loaded dry casks, half loaded dry casks, dry casks with one row of fuel assemblies missing, dry casks with one fuel assembly missing and empty dry casks. Various detector sizes (1.2 m x 1.2 m, 2.4 m x 2.4 m and 3.6 m x 3.6 m) and number of muons ($10^5$, $5 \cdot 10^5$, $10^6$ and $10^7$) were used to assess the effect on image resolution. The Point-of-Closest-Approach (PoCA) algorithm was used for the reconstruction of the stored contents. The results demonstrate that multiple Coulomb scattering can be used to successfully reconstruct the dry cask contents and allow identification of all scenarios with the exception of one fuel assembly missing. In this case, an indication exists that a fuel assembly is not present; however the resolution of the imaging algorithm was not enough to identify exact location.

*Corresponding author: schatzid@purdue.edu

## I. INTRODUCTION

Since the pioneering work of E.P. George [1] and L. Alvarez [2], relativistic muons have been shown to have the ability to penetrate dense materials and by monitoring the subsequent scattering and/or attenuation of muons, a measurable signal about the structure and composition of the interrogated material can be obtained [3]. Recently, cosmic ray muons have been investigated for volcano imaging and cargo scanning applications and their use has been extended to nuclear waste imaging and determination of molten nuclear fuel location in nuclear reactors having suffered from the effects of a severe accident similar to the one happened in Fukushima [4-23]. Earlier muon radiographic techniques were based on attenuation principles. A new promising method based on multiple Coulomb scattering was developed and demonstrated at LANL for detection of high-Z materials hidden in a large volume of low-Z materials, a situation representative of shielded material shielded hidden in a cargo container [24, 25]. It was suggested that muon momentum measurement which can be achieved by indirectly measuring muon scattering in several layers of materials with known thickness could improve image resolution. This was later demonstrated in the CRIPT prototype developed for cargo scanning applications [17]. Recent efforts focus on muon radiography for location and material identification of radioactive objects within closed containers. Notable examples are experimental and simulation efforts to characterize legacy nuclear waste containers [18, 19], reconstruct the contents of concrete containers hosting CANDU type spent nuclear fuel [20] and obtain information from copper canisters storing spent nuclear fuel [21, 22]. Image reconstruction of test objects was initially performed using a geometry-based reconstruction algorithm, known as the Point-of-Closest-Approach (PoCA) [25] inspired in principle by earlier work on nuclear scattering radiography [26]. Advanced statistical reconstruction techniques based on information obtained from muon scattering and displacement improved image reconstruction at the expense of computation and memory usage [27]. Computational performance was improved through updated versions using expectation maximization principles [7]. Alternative techniques to imaging have been proposed and include the development of monitoring algorithms suitable to provide a fast yes/no decision [28-32]. Existing muon detectors use, among others, drift-wire chambers, e.g., the Large and Mini Muon Trackers developed at LANL [33], and scintillating fibers, e.g., the CRIPT detector developed at AECL [17]. Delay line chambers have also been used at LANL [5] and small prototypes with gas electron multiplier detectors have been developed by the HEP lab at

Florida Tech to determine muon positional information and demonstrate the application [27]. Resistive plate chambers are also explored as a cost effective alternative to muon detection [34]. Cox et al. summarized the requirements for muon detector development concluding that coincidence timing in the order of nanoseconds, spatial resolution in the order of sub-mm and energy determination will be required for future applications [35].

The role of cosmic ray muons becomes especially important since their use can be extended to non-destructive assessment of nuclear materials stored within sealed dense containers [9]. Muons present significant advantages over the existing monitoring and imaging techniques. Among others, utilization of muons requires no radiological sources and consequently the absence of any artificial radiological dose. Controlling nuclear material at its source is one of the main strategies to minimize the risks of nuclear proliferation and reduce potential homeland threats. Since the early 1950's, when the first nuclear power plant began to produce electricity, approximately 65,000 metric tons of spent nuclear fuel have been generated, 25% of which are under dry storage conditions [3]. Visual inspection of a dry cask is not easily achievable after the spent nuclear fuel has been placed inside the dry cask. Conventional methods for examining the interior of materials e.g., x-rays, are limited by the fact that they cannot penetrate very dense well-shielded objects [4] while more sophisticated techniques such as the penetrating neutrons or the recently developed proton radiography necessitate the use of an expensive accelerator [3].

Previous efforts on spent nuclear fuel imaging have used only muon attenuation [23] or focused on obtaining signals from copper canisters without, however, any image reconstruction [21, 22]. In the present work, the applicability of multiple Coulomb scattering of muons for imaging spent nuclear fuel in commercial NRC-approved vertical and horizontal dry casks is investigated. The goal is to use muons to differentiate between spent nuclear fuel dry casks and reconstruct images of the stored contents. The strong dependence of scattering on atomic number Z and density of the material indicate that multiple Coulomb scattering could be an excellent candidate for spent nuclear fuel imaging. Calculations of muon scattering were performed for various scenarios, including vertical and horizontal fully loaded dry casks, half loaded dry casks, dry casks with one row of fuel assemblies missing, dry casks with one fuel assembly missing and empty dry casks. The actual muon energy spectrum and angular distribution were reproduced using a "Muon Event Generator" [37]. The "Muon Event Generator" generates muons that follow the experimentally measured seal level spectrum. The "Muon Event Generator" was coupled with the Monte Carlo code GEANT4 [38] and the passage of muons through a commercial dry cask loaded with PWR fuel assemblies was simulated. To determine the sensitivity of the technique to perform imaging of spent nuclear fuel dry casks, simulations were performed with $10^5$, $5 \cdot 10^5$, $10^6$ and $10^7$ muons and muon detector sizes having active area 1.2 m x 1.2 m (small), 2.4 x 2.4 m (medium) and, 3.6 x 3.6 m (large). The detectors are idealized gas chambers filled with argon gas and can record muon position, and direction Momentum measurement was not considered. The obtained accumulated dataset of muon scattering angles is processed with the PoCA imaging algorithm.

## II. DRY CASK SIMULATIONS

*A. Muon-dry cask interactions*

As a result of the processes occurring in the atmosphere, a cascade of secondary rays and relativistic particles is created. Among these particles, pions and kaons decay almost instantaneously to muons and give rise to a considerable muon flux that reaches sea level. Cosmic ray muons are unstable particles of two charge types, having approximately 200 times the mass of electron, generated naturally in the atmosphere, and rain down upon the earth at an approximate rate of 10,000 particles $m^{-2}$ $min^{-1}$ [39]. The muon spectrum has been experimentally measured and shown to vary significantly with energy and zenith angle. The experiments cover a wide range of energies, from 0.2 to 20,000 GeV, zenith angles from from $0^o$ to $89^o$ and 10 m to 1270 m altitude [40]. Above 10 GeV, the spectrum follows a power law profile which is practically independent of the zenith angle. For energies less than 10 GeV, the spectrum slowly increases until a maximum value at ~0.5 GeV and then decreases rapidly. For less than 10 GeV the zenith angle dependency increases significantly. It is important to accurately predict the muon spectrum as it can have a great impact in the design of experiments for muon applications and the design of shielding geometries for neutrino experiments.

Relativistic muons are minimum ionization particles and interact with matter in two primary ways, energy loss from inelastic collisions with electrons and angular deflection from nuclei due to multiple Coulomb scattering. Energy loss via Bremsstrahlung emission is negligible due to the large muon mass. The overall deflection of a muon exiting a material is the sum of all the small random contributions during a muon's path [41]. Moliere derived the angular scattering distribution which has been shown to be in good agreement with most of experimental data.

Moliere started from Rutherford's single scattering formula using the small angle approximation and the Thomas-Fermi model of the atom to account for the screening of the Coulomb potential by the atomic electrons. Moliere's scattering distribution can be described approximately as a Gaussian distribution where the root mean square (RMS) scattering angle of the Gaussian part in is given by [42]:

$$\theta_{RMS}^{M} = \chi_c \sqrt{B}, \tag{1}$$

where:

$$B = 1.153 + 2.583 \log(\chi_c/\chi_a)^2. \tag{2}$$

Moliere defined a characteristic angle, $\chi_c$, which was later corrected by Bethe to account for the inelastic scattering off atomic electrons. For a homogeneous target it is given by:

$$\chi_c^2 = \frac{4\pi N_A \rho X z^2 Z(Z+1) e^4}{A p^2 \beta^2 c}, \tag{3}$$

where $N_A$ is Avogadro's number, Z and A the atomic and mass number, ρ is the density of the materials, $X$ material thickness. The Thomas-Fermi model of the atom was used to account for the screening of the Coulomb potential by the atomic electrons and the contribution is described by an expression for the screening angle, $\chi_a$:

$$\chi_a^2 = \chi_0^2 (R + 3.76\eta^2), \tag{4}$$

where R=1.13 is constant for all Z and:

$$\chi_0 = 1.13\alpha (m_e c/p) Z^{1/3}, \tag{5}$$

$$\eta = \frac{zZe^2}{\hbar v}. \tag{6}$$

Simulations with $10^6$ muons with energy in the range 1-60 GeV result in scattering variance estimates for a vertical fully loaded dry cask, a half loaded and an empty one are 4119.453, 2792.103 and 1493.074 mrad², respectively. The scattering variance distribution is shown to follow Gamma distribution and its moments can be derived as shown in [30]. Then, using Chebyshev's inequality, it can be shown that the scattering variance estimates are within 1% of the true (theoretical) value at a 99% confidence level.

*B. Dry cask model*

A variety of dry cask designs have been produced for use in storage installations over the years which slightly vary in design and characteristics depending on cask manufacturer. These can be classified into two main categories, i) metal canisters shielded within vertical concrete or metal overpack and, ii) metal canisters shielded in horizontal concrete modules [43]. Gamma shielding is provided by the cask itself, also known as overpack.

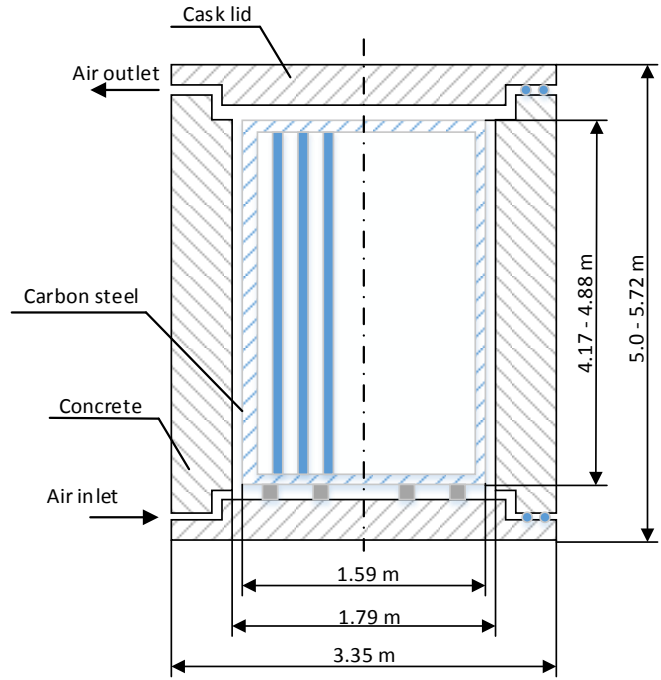

Fig. 1. Dimensions and components of a typical dry cask [43].

The fuel assemblies are stored within a metallic basket located inside the canister which consists of welded borated stainless steel plates arranged in square channels. Typically, commercial dry casks provide space for 24-32 PWR or 56-86 BWR assemblies [43]. They can be stored either vertically or horizontally and weigh more than 100 tons when loaded (fuel assemblies weigh approximately 0.5 – 1 Metric Ton of Uranium each). Fig. 1 shows the basic dimensions and components of a typical commercial dry cask.

In the present work, the GEANT4 Monte Carlo code [38] was used for simulating muons and their paths through spent nuclear fuel dry casks. GEANT4 is a Monte Carlo code designed and developed for use by the High Energy Physics community for the tracking of subatomic particles and their interactions with matter. GEANT4 simulations require knowledge of the muon spectrum and angular distribution and the capability to repeatedly generate random samples from these distributions. GEANT4 does not provide a built-in library for muon energy and angular distribution. To overcome this difficulty, a "Muon Event Generator", i.e., a cosmic ray muon sampling method, using a phenomenological model that captures the main characteristics of the experimentally measured spectrum coupled with a set of statistical algorithms, is used [37]. Dry cask models consist of vertically and horizontally oriented commercial dry casks, loaded with 24 PWR or less fuel assemblies with a concrete overpack. A representative dry cask having average characteristics of currently available commercial dry casks was selected. External cask dimensions are 3.658 m height and 3.37 m outer diameter. The dry cask has concrete (density 2.3 g/cm$^3$) walls, 3.658 m high and with inner and outer radius of 0.8635 m and 1.685 m, respectively. Fuel assemblies are rectangular in shape, 214.5x214.5 mm, and comprise of 15x15 $UO_2$ fuel rods, 3.658 m high and 10.7 mm in diameter. Several assumptions have been made to simplify and speed up the simulations: canister structure and inner basket have been neglected due to their small thickness, fuel cladding is also neglected as it is not expected to significantly influence muons due to its very small thickness and plutonium and fission products have not been taken into account as they represent a very small fraction (~3%) compared to uranium in the fuel assemblies. Preliminary investigations indicated that inclusion of canister and fuel basket structure, although important for noise studies and detector design, have no observed effect on the reconstructed image. Figs 2 and 3 depict the GEANT4 model of fully loaded vertical and horizontal dry casks.

Muon detection and scattering measurement requires placement of detectors on two opposite sides of the object to be inspected. The detectors used in the present work are idealized gas chambers, 1 cm thick, filled with argon gas at standard temperature and pressure and can record muon position, and direction. The detectors were assumed to have perfect resolution. Only the muon component of cosmic rays was simulated, other secondary particles were neglected. The muon source covers the detector active area and is initiated at a distance 20 cm from the first detector plane. The detectors are modeled as parallel planes of position sensitive chambers, e.g., drift-wire chambers or gas-electron multiplier detectors, which record the position of each muon before and after interaction with the dry cask. Each muon will result in four recorded position measurements. Only muons that pass through all detector planes were recorded. Cosmic ray muons pass through the first detector plane, i.e., farthest detector plane through which a muon passes, and their initial positions are recorded. The muons then pass through concrete and $UO_2$, exiting through another layer of concrete before hitting and interacting with the second plane of detectors where their final positions are also recorded. This information is then processed using imaging algorithms.

Fig. 4 shows the simulated scenarios. To determine the sensitivity of the technique for imaging spent nuclear fuel dry casks simulations performed for $10^5$ to $10^7$ muons and included three detector sizes with active areas 1.2 m x 1.2 m (small) 2.4 m x 2.4 m (medium) and 3.6 m x 3.6 m (large). The amount of muons passing through the detector planes depends on the size, orientation and detector solid angle. The angular distribution can be approximated by [44]:

$$\frac{dN}{d\Omega} = \frac{3}{\pi} \cos^2 \theta, \qquad (7)$$

where θ is the zenith angle. The detector solid angle is:

$$\Omega = \frac{hw \sin \theta}{d^2}, \qquad (8)$$

where h and w represent the dimensions (width, height) of the detector planes and, d is the distance from the dry cask centerline to the first detector plane. The number of muons passing through a detector is:

$$N = \frac{dN}{d\Omega} \Omega \, hw \sin \theta . \qquad (9)$$

A horizontal detector located at zenith angle 89°, with dimensions h=w=3.6 m and d=2.5 m would require approximately 1 and 10 days to gather $10^5$ and $10^6$ muon measurements, respectively. This result does not take into account other factors that would increase measurement time, for example background radiation field and detector noise.

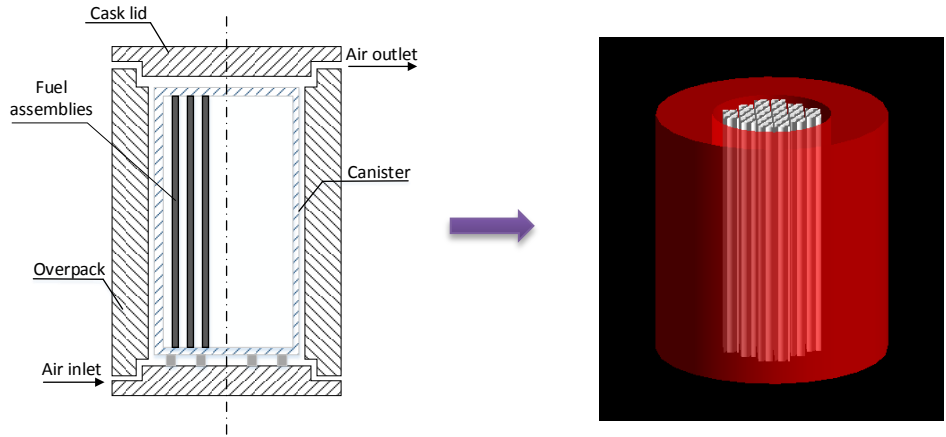

Fig. 2. GEANT4 model of a vertical fully loaded dry cask.

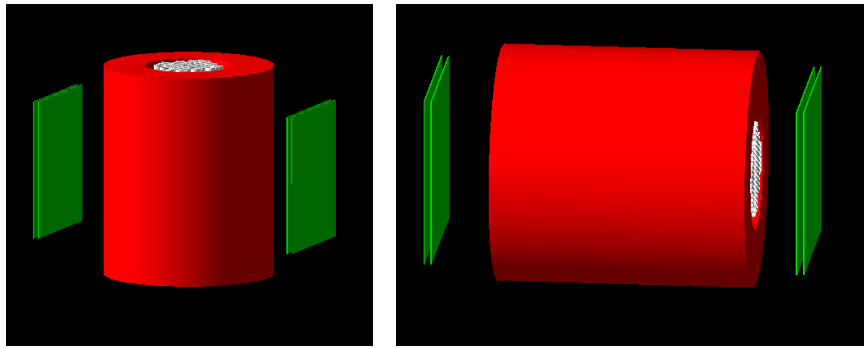

Fig. 3. GEANT4 model of a vertical fully loaded dry cask and associated muon detectors (left) - GEANT4 model of a horizontal fully loaded dry cask and associated muon detectors (right).

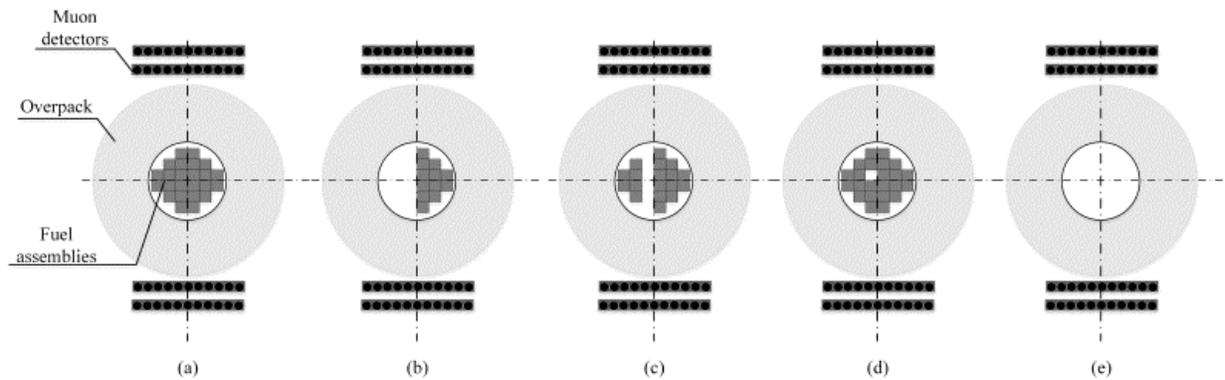

Fig. 4. Simulated dry cask scenarios; (a) fully loaded, (b) half loaded, (c) row of fuel assemblies missing, (d) one fuel assembly missing and, (e) empty.

*C. Code Benchmarking*

Prior to muon-dry cask simulations, the use of GEANT4 to simulate the passage of muons through a test object was validated. As no experimental results of dry casks are currently available, the first muon scattering demonstration by LANL was used as a benchmark. Purpose was to verify proper use of GEANT4, provide confidence on correct modeling of the muon detectors and the capability of the imaging algorithm to correctly reproduce experimental measurements. The LANL muon prototype consisted of four wire chambers with active area 60x60 cm$^2$, filled with a gas mixture 65% argon and 35% isobutene, assembled into a stack and separated vertically by 27 cm. The detectors were triggered by plastic scintillators and resolution of the apparatus was found to be 400 μm FWHM. A tungsten cylinder was used as a test object. The cylinder, 11 cm in diameter and 5.7 cm in height, was placed on a plastic platform supported by two steel beams of 3 mm thickness. Fig. 4a shows the detector prototype, the tungsten object and the steel beams. More details on the experiment setup can be found in [5]. The geometry of the detectors, the tungsten object and the steel beams were modelled in GEANT4. The muon spectrum and angular distribution were reproduced using the "Muon Generator". The muon source was a plane located above the muon detectors with area twice that of the detectors. The GEANT4 model of the detectors and muon trajectories are shown in Figs 4b and 4c. 100,000 muons were recorded and the PoCA reconstruction method was used. A voxel size of 1 cm$^3$ and an object volume of 60 cm x 60 cm x 30 cm were used. The LANL experimental apparatus with tungsten object and steel beams is shown in Fig. 5. Experimental and simulated results appear in Fig. 6. The simulated reconstructed images are very similar to the experiment. The shape of the tungsten cylinder is correctly reconstructed and the steel beams appear as well.

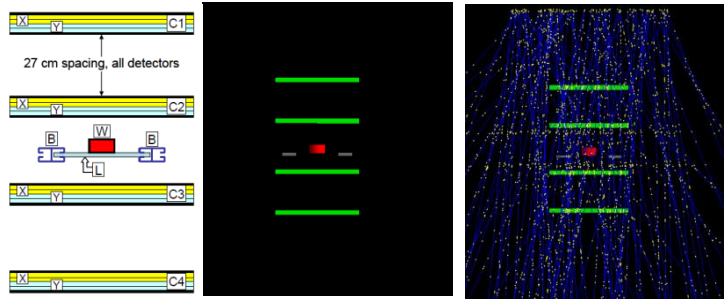

Fig. 5. LANL experimental apparatus with tungsten object and steel beams (Reprinted by permission from Ref. 5) (left) - GEANT4 model (center) - Muon source and muon trajectories (right).

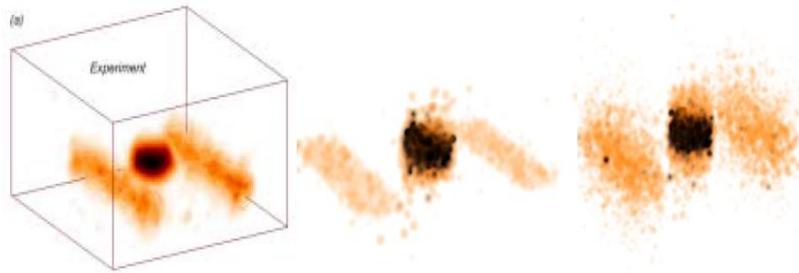

Fig. 6. Experiment (Reprinted by permission from Ref. 5) (left) - Simulation without noise (center) - Simulation with 400 μm FHWM Gaussian noise (right).

## III. DRY CASK IMAGING

The simulated incoming and outgoing positions of muons are recorded and processed to compute the scattering angle for each muon. The accumulated dataset of muon scattering angles is used as input to the PoCA imaging algorithm. The algorithm calculates the point of closest approach between the incoming and outgoing trajectories which in 2D is the intersection of the two lines. For the 3D case, in which the trajectories may not intersect, the algorithm calculates the minimum distance between the incoming and outgoing muon trajectories and assigns the scattering event to the middle point of that distance to a voxel on a 3D grid. The algorithm depends on the number of muons passing through the object volume and presents computational complexity $O(N)$ [27]. The algorithm does not take into account the physics of multiple Coulomb scattering; instead, it makes the assumption of a single scattering event. This assumption presents certain limitations to the resolution of the algorithm. Although algebraic reconstruction techniques may have better resolution are far more computationally demanding than PoCA. PoCA is a relatively fast algorithm compared to other reconstruction algorithms especially when reconstructing large geometries with hundreds of thousands of voxels, e.g., dry casks, within a reasonable amount of time. Efforts to use the EM algorithm [7] resulted in running times so slow with the available computing resources that would make the reconstruction of all the scenarios under investigation prohibitively long.

During the reconstruction, each pixel in 2D or voxel in 3D contains the accumulated scattering angles of muons passing through that point. It is expected that materials with high-Z will appear with higher values in the reconstructed images and voxels with high values would represent high-Z materials where muons scattered significantly. Let $L_1$ the incoming trajectory and $L_2$ the outgoing trajectory. The algorithm first calculates the minimum distance between the two lines, described mathematically as follows:

$$dist(L_1, L_2) = \min dist(P, Q), \qquad (10)$$

where P, Q are points on the incoming trajectory $L_1$ and the outgoing trajectory $L_2$, respectively. Applying analytical geometry principles, we find two points where the minimum distance between the two lines occurs:

$$d = dist(L_1, L_2) = |P(s_c) - Q(t_c)| \qquad (11)$$

The two lines can be written in 2-dimensional space:

$$L_1 := P(s) = P_0 + s(P_1 - P_0) = P_0 + s\boldsymbol{u} \qquad (12)$$

$$L_2 := Q(s) = Q_0 + t(Q_1 - Q_0) = Q_0 + t\boldsymbol{v} \qquad (13)$$

The minimum distance is then given by the following expression:

$$d = \left|(P_0 - Q_0) + \frac{(be - cd)\boldsymbol{u} - (ae - bd)\boldsymbol{v}}{ac - b^2}\right| \qquad (14)$$

where $a=\boldsymbol{u}\cdot\boldsymbol{u}$, $b=\boldsymbol{u}\cdot\boldsymbol{v}$, $c=\boldsymbol{v}\cdot\boldsymbol{v}$, $d=\boldsymbol{u}\cdot\boldsymbol{w_0}$, $e=\boldsymbol{v}\cdot\boldsymbol{w_0}$, and $\boldsymbol{w_0}=P_0-Q_0$. The point of the closest approach is located midway through the minimum distance and is associated with a single pixel on a 2D grid or a voxel on a 3D grid. At the estimated point of scatter the following signal value is assigned (Fig. 7):

$$s_i = \frac{1}{2}\left[\left(\theta_x^{out} - \theta_x^{in}\right)^2 + \left(\theta_y^{out} - \theta_y^{in}\right)^2\right] \quad (15)$$

As more muons pass through the object volume, each pixel (or voxel) is filled with scattering angles. The final value of each voxel is the average scattering angle divided by the voxel size L:

$$Voxel\ value = \frac{1}{NL}\sum_{i=1}^{N} s_i \quad (16)$$

Investigations with median value instead of average scattering angle showed little to no improvement on the reconstructued images. This can be attributed to the large voxel size used thus the smoothing effect of the median has little effect on resolution.

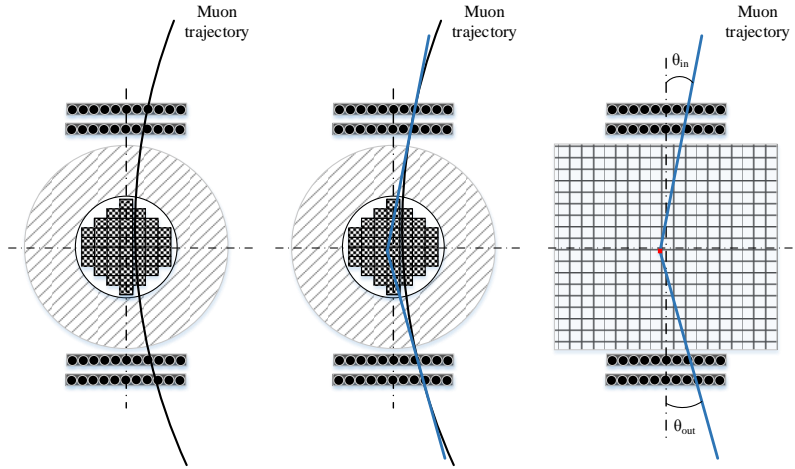

Fig. 7. PoCA algorithm principle (left: muon trajectory through muon detectors, spent nuclear fuel and dry cask, center: extrapolated linear incoming and outgoing trajectories, right: superimposed grid and assigned scattering point)

## IV. RESULTS & DISCUSSION

The PoCA algorithm was used for the reconstruction of vertical and horizontal dry casks loaded with different amount of spent nuclear fuel. Three detector sizes (small, medium, large) were used to investigate the effect of detector size on image resolution. Additional simulations and image reconstructions were performed for $10^5$, $5 \cdot 10^5$, $10^6$ and $10^7$ muons. Detector size and number of muons eventually determine the measurement time needed to provide meaningful information about the cask content. For example, large detectors would be preferable since reduced time would be required to obtain the necessary number of muons. On the other hand, large detectors would be more expensive and more difficult to transport on site. For all plots shown, Y-axis is oriented vertically, X-axis is oriented parallel to the detector planes and Z-axis is vertical to the detector planes.

Reconstruction images for vertical and horizontal fully loaded dry casks using $10^5$, $5 \cdot 10^5$ and $10^6$ muons are shown in in the upper and lower row of Fig. 8, respectively. The reconstructions were performed for a large detector (3.6 m x 3.6 m) with muons having energies 1-60 GeV, grid dimension 40x40x40 $cm^3$ and voxel size 10 cm. Fig. 8 shows 2-D slices taken at the x-y plane through the center of the cask. The color scale represents the magnitude of the scattering angles in $mrad^2$/cm. It is observed that even with as low as $10^5$ muons the shape of the cask, the concrete walls and the location of spent nuclear fuel can be identified fairly accurately. Increasing the number of muons, the resolution improves considerably and the location and shape of different materials is shown correctly. Although the location of fuel assemblies is correctly identified, the air gap that exists between the concrete walls and the fuel assemblies is not correctly captured and this is independent of the number of muons. When multiple layers of materials are stacked vertically to the muon trajectory, the algorithm assigns scattering points between the layers and points are placed in areas where there are no fuel assemblies. This can be attributed to the PoCA algorithm which assigns scattering points in between multilayered materials and the fact that muon tracks are constrained by the acceptance of the detectors resulting in scattering points to tend to be distributed along the axis of geometric acceptance. For horizontal dry casks the resolution appears to be somewhat improved and allows identification of the air gap between concrete and fuel assemblies.

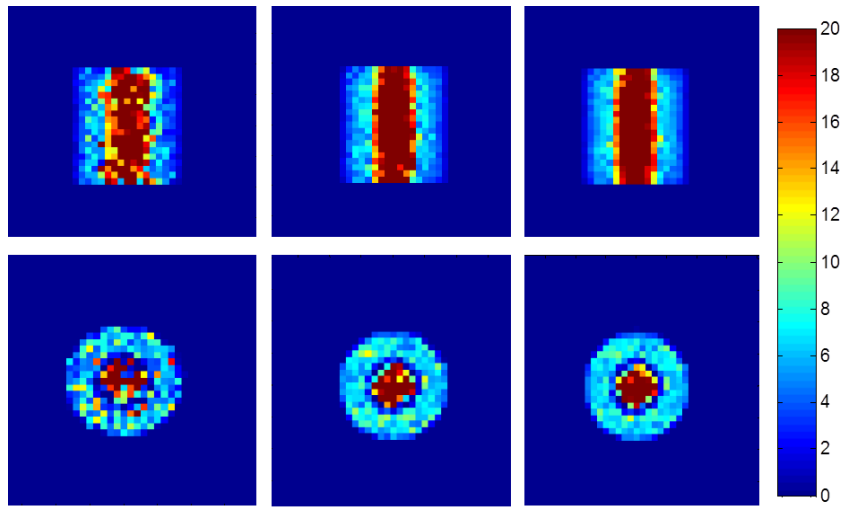

Fig. 8. Imaging results of fully loaded dry casks with different number of muons (upper row: vertical fully loaded dry cask, lower row: horizontal fully loaded dry cask, left column: $10^5$ muons, center column: $5 \cdot 10^5$ muons, right column: $10^6$ muons). Scale in mrad$^2$/cm.

In Figs 8 and 9, reconstructed images are shown for fully loaded and empty vertical and horizontal dry casks, respectively. The cask shape is reconstructed correctly and concrete walls can be identified in all cases. Comparison between Figs 8 and 9, demonstrates that a fully loaded dry cask can be differentiated against an empty one using as low as $10^4$ muons. However, as it will be shown below, the resolution is too low to allow identification of more complex cases. The fully loaded and empty dry casks can be used as a baseline to compare against other scenarios. Reconstructions of vertical fully loaded, half loaded and empty dry casks generated using $10^6$ muons and different detector sizes are shown in Fig. 10. The reconstructions using small size detectors are shown in the upper row of Fig. 10. Reconstructed images using medium and large sized detectors are shown in the middle and lower row of Fig. 10, respectively. The algorithm can reproduce the cylindrical shape and allow differentiation between the concrete walls and the fuel. For the empty dry cask, the center is filled with air which the algorithm incorrectly interprets as concrete. Reconstructed images of horizontal fully loaded, half loaded and empty dry casks are shown in Fig. 11. The PoCA limitation is reduced for horizontal dry casks. In this case, both fully loaded and empty dry casks are correctly reconstructed and better resolution is observed for horizontal dry casks than vertical ones. The reconstructed images allow correct identification of the dry cask loading and separation between concrete and fuel assemblies.

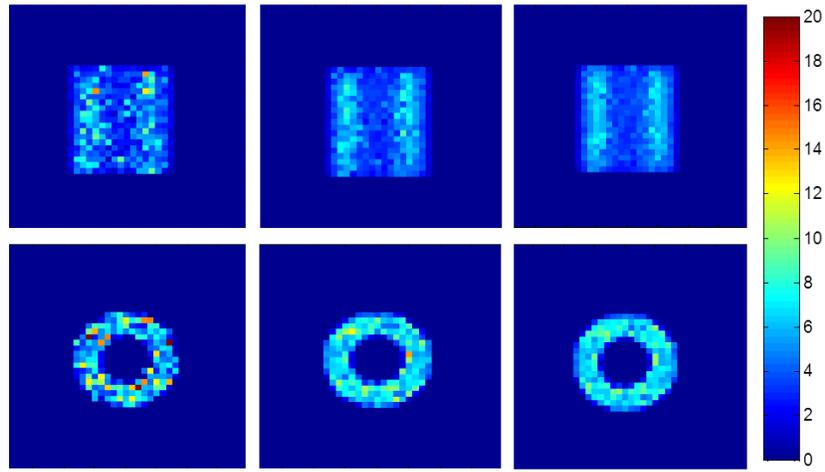

Fig. 9. Imaging results of empty dry casks with different number of muons (upper row: vertical empty dry cask, lower row: horizontal empty dry cask, left column: $10^5$ muons, center column: $5 \cdot 10^5$ muons, right column: $10^6$ muons). Scale in mrad$^2$/cm.

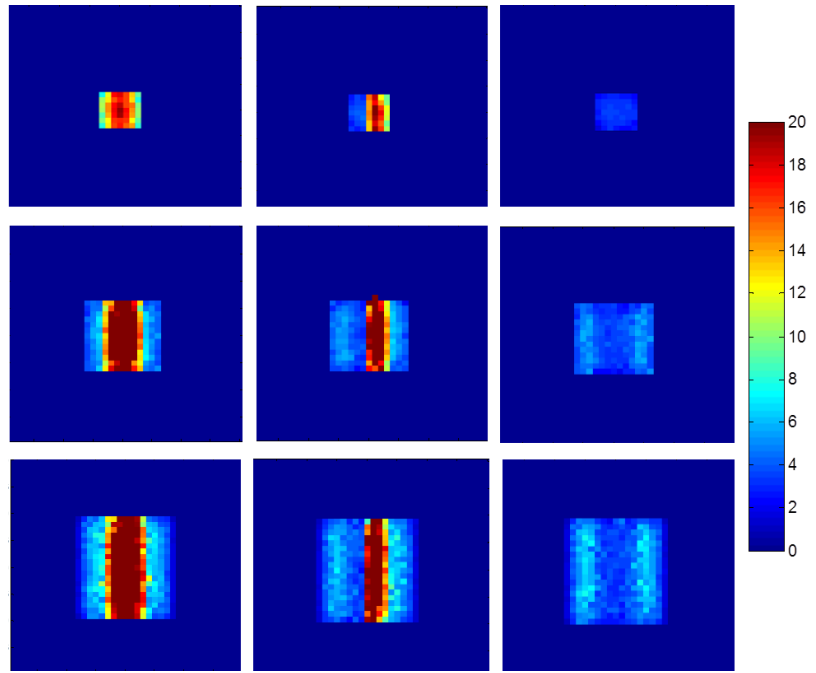

Fig. 10. Imaging results of vertical dry casks with different muon detector sizes (upper row: small size detector, middle row: medium size detector, lower row: large size detector, left column: fully loaded, center column: half loaded, right column: empty). Scale in mrad$^2$/cm.

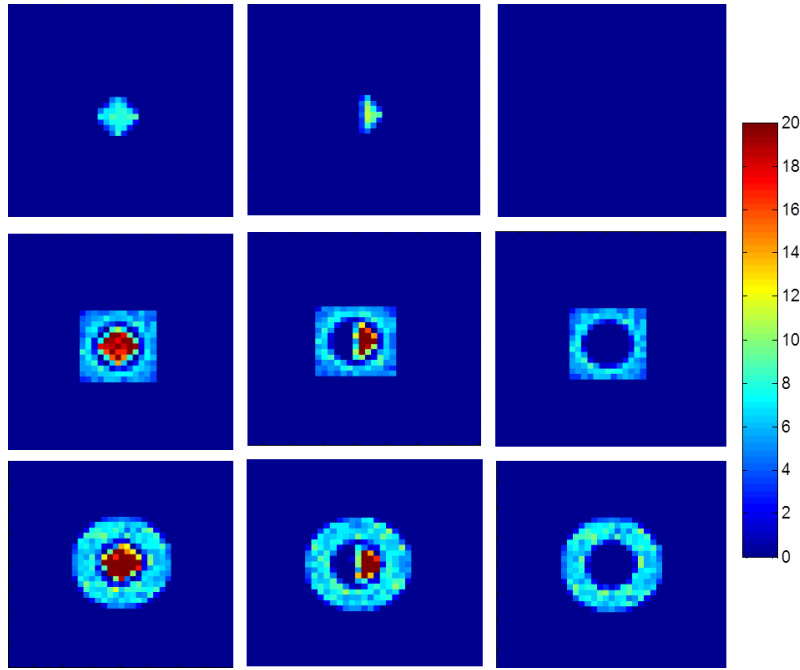

Fig. 11. Imaging results of horizontal dry casks with different muon detector sizes (upper row: small size detector, middle row: medium size detector, lower row: large size detector, left column: fully loaded, center column: half loaded, right column: empty). Scale in mrad$^2$/cm.

Image reconstructions were quantized to investigate the possibility of material classification. The accumulated scattering angles were classified into three categories: low (black color), medium (grey color) and high density (white color) materials. Scattering of 20 mrad$^2$/cm or greater is classified as high density material, medium density is shown by scattering in the range 10-20 mrad$^2$/cm and low density material by scattering of 10 mrad$^2$/cm. The results are shown in Fig. 12 for vertical fully loaded, half loaded and empty dry casks and Fig. 13 for horizontal fully loaded, half loaded and empty dry casks. The reconstructed images are compared against the reconstructed images that would result from an ideal algorithm with perfect resolution. The processed reconstructed images facilitate differentiation between materials and identification of concrete and fuel assembly location. Air (black color) and concrete (grey color) can be easily separated over fuel assemblies (white color). The algorithm correctly identifies the amount of loading for each of the three cases (fully loaded, half loaded and empty) but misidentifies air as concrete in the case of an empty cask. This limitation is reduced for horizontal geometries where the empty dry cask is correctly classified as having no fuel within.

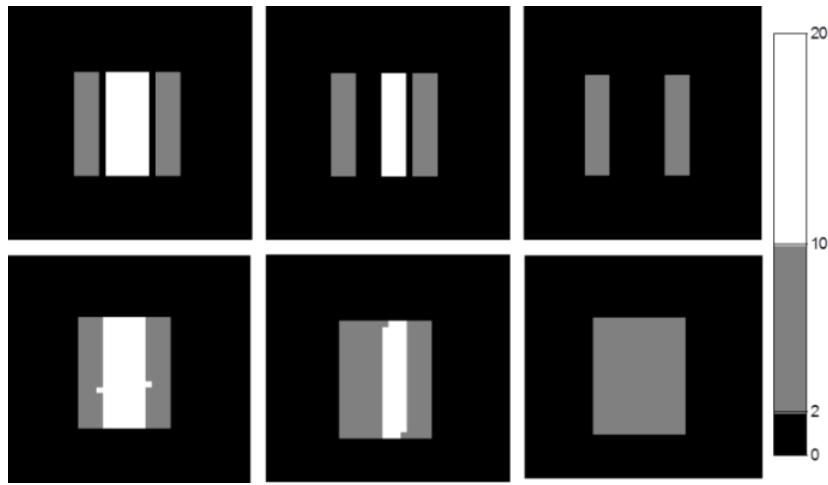

Fig. 12. Imaging results of a vertical dry cask (upper row: images using a perfect algorithm, lower row: images using PoCA, left column: fully loaded, center column: half loaded, right column: empty). Scale in mrad$^2$/cm.

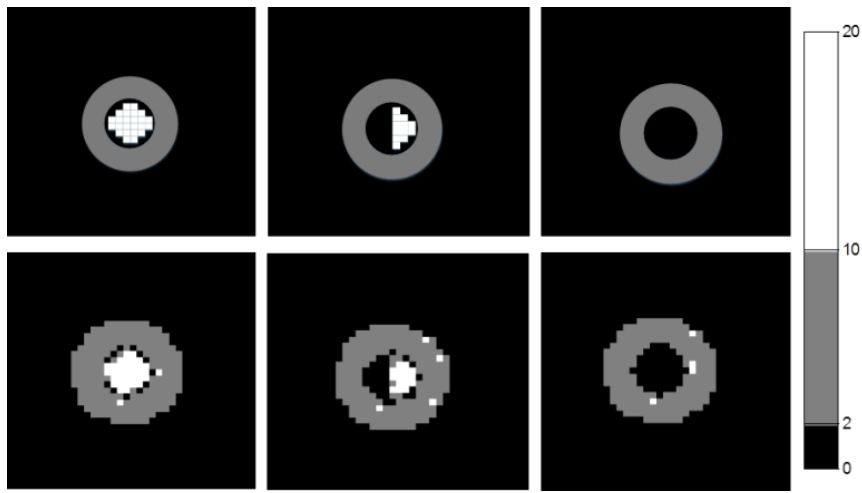

Fig. 13. Imaging results of a horizontal dry cask (upper row: images using a perfect algorithm, lower row: images using PoCA, left column: fully loaded, center column: half loaded, right column: empty). Scale in mrad$^2$/cm.

It was shown that reconstructions using the PoCA algorithm allow fairly accurate differentiation among fully loaded, half loaded and empty dry casks with minimal processing and making use of only one 2-D slice of the 3-D object volume. More information would be needed to allow adequate identification in more complex scenarios where smaller fuel quantities are not present. To assess such scenarios, simulations were performed for dry casks where one row of fuel assemblies was missing. Results are shown in Figs 14 and 15. Although one 2-D slice can provide information related to missing fuel assemblies, it is not enough to obtain exact location and number of assemblies missing. Using x-y and x-z slices, the missing row of fuel assemblies can be identified with acceptable accuracy. It tis noted that this can be achieved only when enough muons have been collected, in this case more than $10^5$. The reconstructed images were processed using appropriately selected colormap and smoothed to improve visibility and facilitate differentiation of the important dry cask features. Fig. 16 shows the aesthetically enhanced images for horizontal and vertical dry casks. The different scenarios can be identified with the exception of the case where one fuel assembly is missing. For vertical dry casks it is not possible to locate the position of the missing fuel assembly. For horizontal dry cask, an indication exists that a fuel assembly is not present, but resolution is not enough to identify exact location. This can be attributed to the single scattering event assumption inherent within the PoCA algorithm. The average scattering density for vertical dry casks with one row of fuel assemblies missing is 11.28±1.52. Compared with a fully loaded vertical dry cask 18.63±2.45, the identification can be made with a high degree of confidence. Similarly, the average scattering density for horizontal dry casks with one row of fuel assemblies missing is 12.74±2.07 which compared with a fully loaded one, 28.18±2.86, results in correct identification. For one fuel assembly missing in horizontal cask, the average scattering density is 15.21±1.59, decreasing the confidence margin although approximate identification of missing fuel assembly location is still possible. For the vertical case the margin is even smaller. The blurring effect caused by the PoCA increases the difficulty of correctly locating the missing fuel assembly. Higher statistics could improve resolution and missing fuel assembly identification.

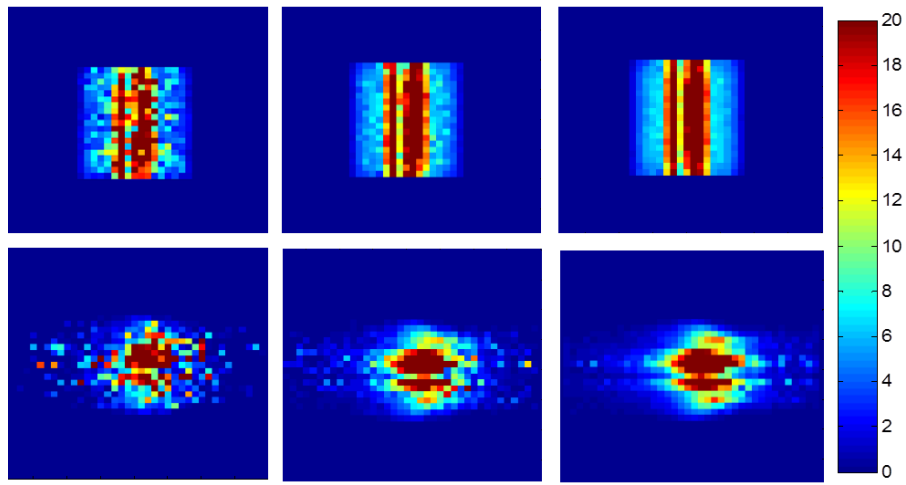

Fig. 14. Imaging results of vertical dry casks with one row of fuel assemblies missing (upper row: vertical slice, lower row: horizontal slice, left column: $10^5$ muons, center column: $10^6$ muons, right column: $10^7$ muons). Scale in mrad$^2$/cm.

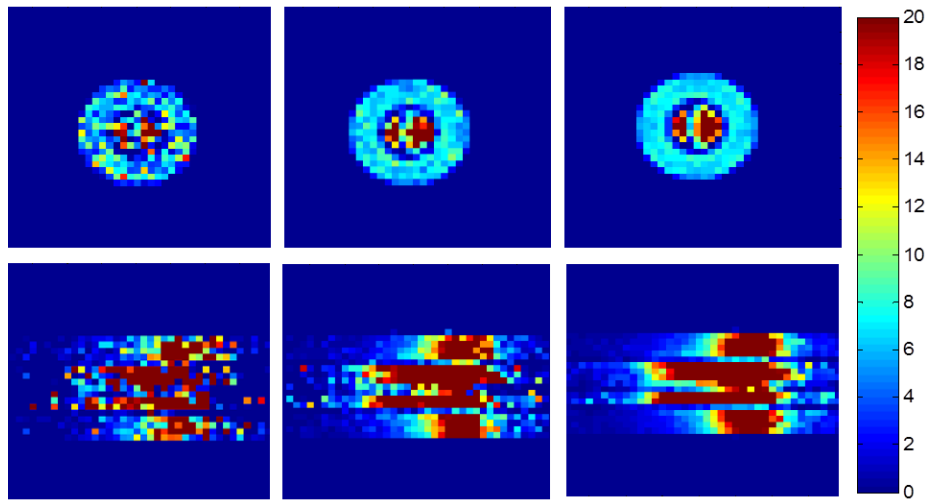

Fig. 15. Imaging results of horizontal dry casks with one row of fuel assemblies missing (upper row: vertical slice, lower row: horizontal slice, left column: $10^5$ muons, center column: $10^6$ muons, right column: $10^7$ muons). Scale in mrad$^2$/cm.

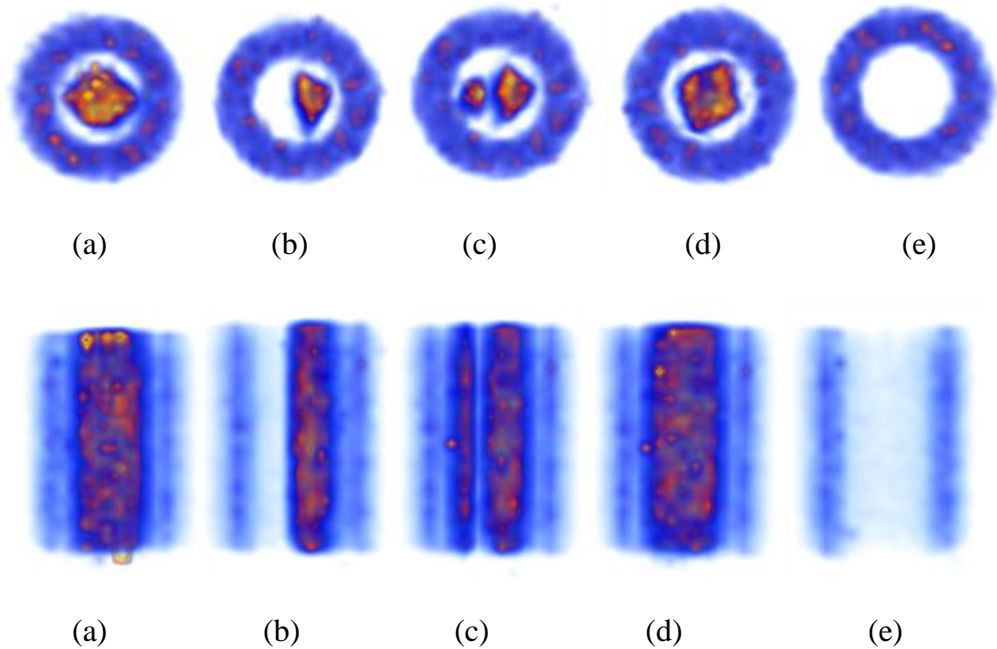

Fig. 16. Imaging results of horizontal (upper row) and vertical (lower row) dry casks processed for aesthetically improved results using heuristics (a: fully loaded, b: half loaded, c: one row is missing, d: one fuel assembly is missing, e: empty).

## V. CONCLUSIONS

Calculations of muon scattering were performed for various scenarios, including vertical and horizontal fully loaded dry casks, half loaded dry casks, dry casks with one row of fuel assemblies missing, dry casks with one fuel assembly missing and empty dry casks. The PoCA algorithm was used for the reconstruction of vertical and horizontal dry casks with different amount of spent nuclear fuel loading. A "Muon Event Generator" was coupled with the Monte Carlo code GEANT4 and muon interactions through a commercial dry cask loaded with PWR fuel assemblies were simulated. Different detector sizes (1.2 m x 1.2 m, 2.4 m x 2.4 m and 3.6 m x 3.6 m) and number of muons ($10^5$ to $10^7$) were used to assess the effect on image resolution. It was shown that reconstructions using the PoCA algorithm allow fairly accurate differentiation among fully loaded, half loaded and empty dry casks with minimal processing and making use of only one 2-D slice of the 3-D object volume. A fully loaded dry cask can be correctly identified and differentiated against an empty one with muons as low as $10^5$. Improved resolution was obtained for horizontal dry casks and for larger number of muons which even allowed identification of the air gap between fuel assemblies and dry cask walls. A missing row of fuel assemblies was identified adequately using two 2-D slices and when enough muons have been collected, in this case more than $10^5$. Multiple Coulomb scattering allows identification of all scenarios with the exception of the vertical dry cask case where one fuel assembly is missing. In the horizontal case an indication exists that a fuel assembly is not present, however resolution was not enough to identify exact location.

## ACKNOWLEDGEMENTS

This research is being performed using funding received from the DOE Office of Nuclear Energy's Nuclear Energy University Programs under Contract DE-NE0000695.